\documentstyle[prl,aps,multicol,epsfig]{revtex}
\begin{document}
\draft
\widetext

\title{Quasiparticle bands and superconductivity in bilayer cuprates. }
\author{A.I. Liechtenstein, O. Gunnarsson, O.K. Andersen,
 and R.M. Martin\cite{Urb}}
\address{  Max-Planck-Institut f\"ur Festk\"orperforschung,
Heisenbergstr.1, D-70569 Stuttgart, FRG}
\date{1 September 1995}
\maketitle

\widetext
\begin{abstract}
We analyze the generic features of the energy spectrum
for two coupled CuO$_2$ layers with a realistic extended Hubbard model.
The quasiparticle bands exhibit
flat regions near X(Y) points in the Brillouin zone with  a large
reduction of the bonding-antibonding splitting, and pinning of {\it extended}
van-Hove singularity to the Fermi level, which is more efficient for
a bi-layers than for a single layer. In contrast to the results
with simpler models, the superconducting temperature
for d$_{x^2-y^2}$ pairing is not lowered by the bi-layer hopping.

\end{abstract}

\begin{multicols}{2}

\narrowtext

The role of multiple layers in the high-temperature superconductors
continues to be one of the most intriguing puzzles, with many conflicting
proposals and interpretations.
One proposal is that superconductivity is enhanced by electronic
correlations which are argued to greatly reduce  single-particle hopping
between layers while allowing {\it pair}-tunneling\cite{PWAscience}.
Angle-resolved photoemission (ARPES) studies\cite
{Schen-rev,Campuzano-BIS} of the energy spectrum
near the Fermi energy for bi-layer materials
can in principle answer the questions of the nature of the
bonding and antibonding bands near the Fermi level.  However,
one recent ARPES experiment\cite
{Campuzano-BIS} resolved only one CuO$_2$ band at the
Fermi surface in BSCO-2212,
supporting the idea of a greatly reduced interlayer hopping,
whereas another measurement\cite{Schen-rev} reports two Fermi sheets, in
general agreement with the predictions of band structure calculations.
In addition, there is evidence for important effects of coupling
between the layers in bilayer systems.
 Neutron scattering experiments see maximum
intensity for spin scattering for wave vectors the $q_z\sim \pi /L$, where L
is the interlayer spacing, for both the antiferromagnetic insulator and in
the superconductor\cite{tranq}.

A generic feature which has emerged from the ARPES experiments\cite
{Schen-rev,Campuzano-BIS} is
a ``flat-band'' or ``extended van-Hove singularity'' which is
``pinned'' to the Fermi level for different hole dopings.
Bifurcated saddle points very close to the Fermi level had, in fact, been
predicted by LDA calculations for YBCO-123\cite{Kanazawa} and caused by
dimpling of the CuO$_2$ planes\cite{dimpl}. The flat band observed by ARPES
has also been attributed to many-body
effects\cite{ScalapinoVH,DagottoVH,HankeVH},
and a number of studies have been done for
correlated electronic states in a single CuO$_2$ plane.
The flat region in the quasiparticle spectrum
has been proposed to be a ``fermion condensate''\cite{Khodel,Nozieres},
or a non-Fermi liquid
area formed in (${\bf q},\omega $) space near two-dimensional van-Hove
singularity\cite{DzyalVH}. The one-band\cite{ScalapinoVH,HankeVH} and
three-band\cite{HankeFLEX} Hubbard models as well as t-J model\cite
{DagottoVH} show a flat quasiparticle band just below the Fermi energy,
which has an ``extended'' van-Hove singularity near the X($\pi ,0$)-point,
due to antiferromagnetic spin fluctuations. The same antiferromagnetic
fluctuations at ${\bf q} \approx (\pi ,\pi )$
lead to d$_{x^2-y^2}$ superconductivity\cite
{Scalapino-rev} with a relatively high transition temperature.
For the two-plane Hubbard model\cite{Scalap-PRB,Scalettar}
with a simple tight-binding spectrum a
reduction of superconducting correlations was found due to
the interlayer coupling.

In this paper we study the bi-layer Hubbard model with realistic
LDA derived hopping integrals.
Combination of this realistic tight-binding (TB) model
with many-body effects accounts for
anomalous properties of cuprate superconductors.
Thus in the normal state of our BSCO model,
many body effects strongly reduce the splitting between the
bonding and antibonding bands in the regions in the {\bf k}-space
where the one-body splitting is large, leading to flat bands
near the  chemical potential to within an energy of order room temperature.
On the other hand,
in the regions where the many-body effects are small, there is also
little splitting in the one-particle spectrum, due to geometry of the bi-layer
bonding in the cuprates. We analyze the pinning of the chemical potential to
these  van-Hove singularities in the case of a mono- and bi-layer
model for different hole dopings. A calculation of the superconducting
transition temperature shows
robust stability of the d$_{x^2-y^2}$
state in the ``three-dimensional'' bi-layer case.

Let us start with the Hubbard hamiltonian for a bilayer CuO$_2$ model:

$$
H=\sum_{i\alpha j\beta \sigma }t_{i\alpha ,j\beta }c_{i\alpha \sigma
}^{\dagger }c_{j\beta \sigma }+U\sum_{i\alpha }n_{i\alpha \uparrow
}n_{i\alpha \downarrow }-\mu \sum_{i\alpha }n_{i\alpha \sigma }\, ,
$$
here $t_{i\alpha ,j\beta }$ are the hopping integrals which gives the energy
bands $\varepsilon _n({\bf k})$, $i(j)$ are the site indices inside the
plane and $\alpha (\beta )$ are the plane indices, U\ is the on-site Coulomb
repulsion and $\mu $ is the chemical potential. Due to the mirror plane
symmetry for the bilayer\cite{dimpl} we can define bonding ($+$) and
antibonding ($-$) electron bands $\varepsilon _{+}({\bf k})$ and $%
\varepsilon _{-}({\bf k})$, where ${\bf k}$ is a vector in the two dimensional
Brillouin zone. This symmetry also holds for the quasiparticles excitation
spectrum of the full interacting system. For our studies we have used
different schemes
which range from the simplest model (only nearest-neighbor hopping in-plane
and between planes) to realistic four- and eight-band models,
obtained by integrating out the high-energy degrees of freedom
from the full LDA
calculations for cuprate superconductors\cite{dimpl}.

For BSCO bi-layer materials we use a
four-band Hamiltonian\cite{OKAperp},
which includes the standard Cu-d$_{x^2-y^2}$,
and O-p$_x$,p$_y$ orbitals, plus the ``Cu-s'' orbital which has also some
Cu-d$_{z^2}$ character.  The last band is needed because the standard
3-band model
does not give an adequate description of the valence band and
the Fermi surface\cite{OKAperp}.
The Cu-s orbital provide the 2nd and 3rd nearest neighbor ($t^{\prime }$
and $t^{\prime \prime }$) intra-plane hopping integrals, as well as the
hoping between the CuO$_2$ planes in the low-energy Hamiltonian H.
Diagonalization of its first term yields the (simplified) ``LDA bands'':

\begin{eqnarray}
\varepsilon _{_{\pm }}({\bf k})=-2t(\cos k_x+\cos k_y)-4t^{\prime }\cos
k_x\cos k_y -  \nonumber  \\
2t^{\prime \prime }(\cos 2k_x+\cos 2k_y)\mp t_{\perp }(\cos
k_x-\cos k_y)^2/4 \nonumber .
\end{eqnarray}
Note that the interlayer hopping is anisotropic with the maximum splitting
of $\varepsilon _{+}({\bf k})$ and $\varepsilon _{-}({\bf k})$ at the X($\pi
,0$) and Y($0,\pi $) points
 and no splitting in the direction from $\Gamma (0,0)$
to M($\pi ,\pi $). We chose values of the tight-binding parameters for
BSCO: $t=0.5$ eV, $t^{\prime }/t=-0.3$, $t^{\prime \prime }/t^{\prime
}=0.2$ and $t_{\perp }=0.15$ eV,
close to the LDA parameters\cite {OKAperp}.
The total band width is
$8t=4$eV, and the bonding-antibonding splitting at the X-point is
$2t_{\perp }=0.3$ eV. The screened electron-electron interaction
parameter for Cu$_d$ orbitals obtained in the constrained LDA calculations
for cuprate is 7-8 eV \cite{U-LDA}.
Since the Cu$_d$ character at the Fermi surface is approximately 65\%
\cite{OKAperp} the effective U is reduced by $0.65^2$ and we
 used in the present study $U= 3.2$ eV. This is a bit less than the band
width, but general results are not very sensitive to the real $U$%
-values. Note that most many-body Monte-Carlo calculations\cite
{ScalapinoVH,HankeVH,Scalettar}
 use only a simple tight-binding model: single - $t$
parameter and isotropic $t_{\perp }$. We believe that in the cuprate system
both many body effects and the band dispersion play a crucial
role.

We used the conserving fluctuation-exchange (FLEX) approximation\cite{FLEX},
which gives an energy spectrum for CuO$_2$ plane
almost identical to Quantum Monte Carlo results\cite{HankeFLEX}.
We calculate the self-consistent one-electron Green's function in the normal
state:
$$
G_{\pm }({\bf k},\omega _n)=\frac 1{i\omega _n-[\varepsilon _{\pm }({\bf k})%
-\mu ]-\Sigma _{\pm }({\bf k},\omega _n)} ,
$$
where $\Sigma _{\pm }({\bf k},\omega _n)$ are bonding and antibonding
self-energies and $\omega _n=(2n+1)\pi T$ are the fermion Matsubara
frequencies with $n$ an integer and T the system temperature. It is convenient
to introduce the simple
notation: $k=({\bf k},\omega _n)$ and $\sum_k=\frac T{2N}%
\sum_{{\bf k},\omega _n}$, where N is the total number of momentum points.
The straightforward generalization of the FLEX
approximation\cite{FLEX} to the case of a bilayer two-band
Hubbard model gives the following equation for the self-energy:
$$
\Sigma _{\pm }(k)=\sum_q[V_{+}(q)G_{\pm }(k-q)+V_{-}(q)G_{\mp }(k-q)] ,
$$
where the contribution to effective potential
$V_{\pm }(q)=V_{\pm }^s(q)+V_{\pm }^c(q) $
 from spin and charge fluctuation are:
$$
V^s_{\pm}=\frac 32\frac{U^2\chi _{\pm }}{1-U\chi _{\pm }}-V^{dc}_{\pm},\, \, \,
V^c_{\pm}=\frac 12\frac{U^2\chi _{\pm }}{1+U\chi _{\pm }}-V^{dc}_{\pm},
$$
here
$V^{dc}_{\pm}=\frac 12U^2\chi _{\pm }$
is the double counted contribution  and $V_{\pm }(q)$
is defined self-consistently thorough particle-hole susceptibility for
bi-layer:
$$
\chi _{\pm }(q)=-\sum_k[G_{+}(k)G_{\pm }(k+q)
+G_{-}(k)G_{\mp }(k+q)].
$$
We solve the nonlinear integral FLEX-equations using the fast Fourier transform
(FFT) method\cite{Serene} on the discreet mesh of 64$\times $64 momenta in
the two dimensional Brillouin zone and 600-900 Matsubara frequencies with the
cutoff of 20-30 eV (which corresponds to the temperature range
of 80-200K). Analytical continuation on the real axes was done by
Pad\'e  approximation. The calculations have been carried out for
different hole concentrations: $\delta = 0.1 - 0.4$ per plane.

\begin{figure}
\vskip  0cm
\centerline{\epsfig{file=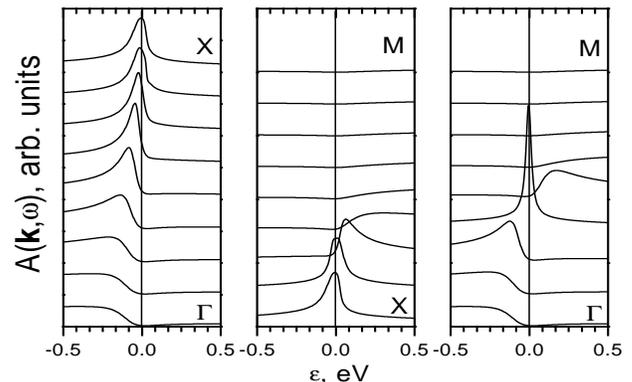,width=8cm,height=5cm}}
\vskip 0.1cm
\caption{The bi-layer spectral function
 for  different
directions in the Brillouin zone ($\delta$ =0.25, T=150 K) .}
\label{Spectral}
\end{figure}

The spectral function $A({\bf k}%
,\omega )=-1/\pi Im(G_{+}({\bf k},\omega )+G_{-}({\bf k},\omega ))$ for
hole doping $\delta =0.25$ per CuO$_2$ and T=150 K
 is shown in
Fig.\ref{Spectral} for $\Gamma $-X-M
directions. One sees that there is only a single peak
with a ``non-Lorentzian'' behavior
crossing the Fermi level.
In fact there are
two bands with the large renormalization of the interlayer splitting from
300 meV to 40 meV,
and there is large broadening at the actual temperature
of the measurement in the normal state.

\begin{figure}
\vskip  0cm
\centerline{\epsfig{file=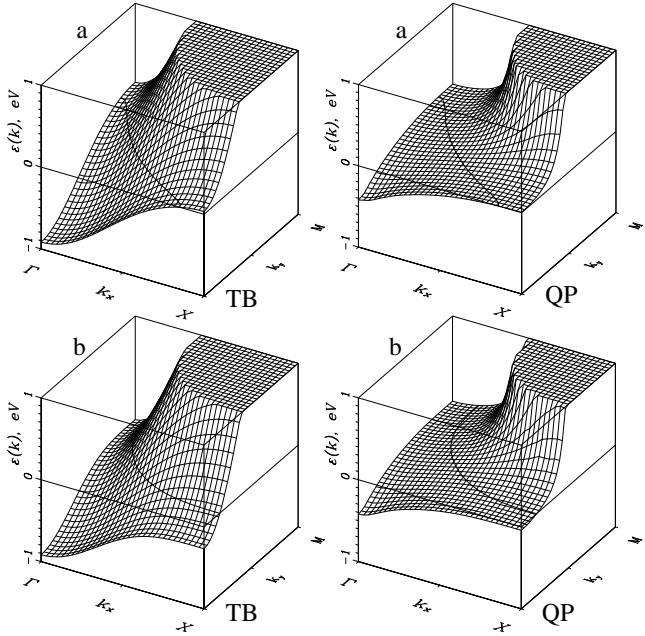,width=8.5cm,height=8.4cm}}
\vskip 0.3cm
\caption{Energy spectrum for U=0 (TB) and U$\neq$0 (QP)
for the antibonding (top) and bonding (bottom) bands.
}
\label{Ek3d}
\end{figure}

This is illustrated in Fig.\ref{Ek3d}, which shows
a comparison of the pure
tight-binding spectrum (U=0) and the quasiparticle spectrum (QP)
for $\delta=0.25$ for the bonding (b) and antibonding (a)
bands together with corresponding Fermi-surface (zero-energy contour).
The quasiparticle dispersion $\varepsilon _{\pm }^{QP}({\bf k})$
have been determined through the
maximum in the spectral function $A_{\pm }({\bf k},\omega )$.
Note that already the TB (or LDA)
bands differ substantially from the simple model with
only one hopping - $t$: the position of the saddle points at the X and Y points
is lower (for ``$t$-model'' they are located at the middle of the band
i.e. at 1 eV in Fig.\ref{Ek3d}). This TB-spectrum changes the
topology of the Fermi surface, and makes the hole doped and the electron doped
situations quite different (as it should be experimentally\cite{Schen-rev})
and suppresses the nesting near half-filling. The last effect increases the
width of a susceptibility peak near M($\pi ,\pi $) point and reduces
the antiferromagnetic spin-correlation length
from $\approx 3$ to $\approx 1$ one lattice spacing.
This leads to the suppression
of the ``shadow bands''\cite{Kampf}
in the quasiparticle spectrum
which occur in the simpler model with
$t^{\prime }=t^{\prime \prime}=0$
and small doping $\delta \approx 0.1 $\cite{QP-Berlin}.

A more important consequence of
$t^{\prime }$ and $t^{\prime \prime }$ is related to
the increased flatness of the van-Hove singularities in the
direction towards $\Gamma $,
which leads to a larger phase space for the spin and
charge excitations. This results in a drastic renormalization of the
energy spectrum (Fig.\ref{Ek3d}), manly due to
spin-fluctuation effects\cite{HankeFLEX}. The quasiparticle bands exhibit
flat regions near X(Y) points in the Brillouin zone with  a large
reduction of the bonding-antibonding splitting.
Note that the susceptibility for anti-ferromagnetic coupling across the
bi-layer $\chi_-(q)$ exceeds that
$\chi_+(q)$,  for ferromagnetic coupling
and near $\vec q = (\pi, \pi)$ the
difference could be even an order of magnitude for $t'=t''=0$, reflecting
the enhancement of the
antiferromagnetic fluctuations between the layer\cite{tranq}.
The Fermi surface is similar to the LDA, but
the large flatting of the
quasiparticle bands and the broadening of the spectral function\cite{Khodel}
is not properly described in LDA.

\begin{figure}
\vskip 0.0cm
\centerline{\epsfig{file=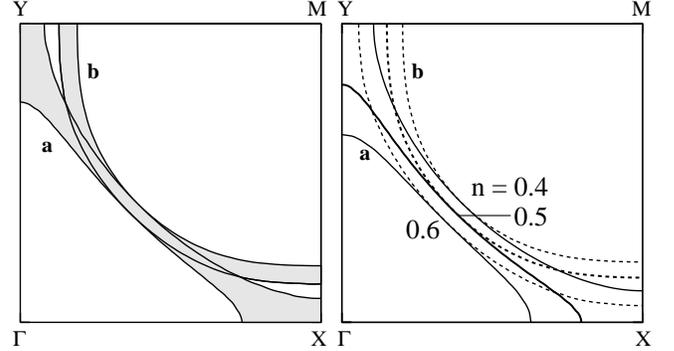,width=8.5cm,height=4.6cm}}
\vskip 0.1cm
\caption{ Fermi surface for $\delta =0.25$, T=150 K:
 temperature broadening (left) and
constant occupation number contours (right).
}
\label{FS}
\end{figure}

We plot in Fig.\ref{FS}
the ``temperature'' smeared Fermi surface ($\mu=\pm T$ for $T=150$K) and
constant occupation number
contours: $n({\bf k})=\sum_nA({\bf k},\omega _n)$ \cite{Randiera}
with $n_0$=0.5
corresponding to the Fermi surface.
Note the extensive regions of nearly equal occupation $\approx 0.5$
near the X and Y points.
It is clear that the difference between the bonding and antibonding
Fermi surface sheets for the BSCO model is very small and hard to detect
experimentally\cite{Schen-rev,Campuzano-BIS}. We believe that the ``shadow
band'' obtained in the recent ARPES
experiment\cite{Campuzano-BIS} is related to
the large dark region near X(Y)-point for the antibonding bands. This flat
region of the quasiparticle spectrum within the room temperature scale from the
Fermi level gives the anomalous linear dependence of the self energy on
temperature and frequency, just from the ``phase-space'' argument\cite
{Nozieres,Gop}, detected numerically near X and Y points\cite
{QP-Berlin}.

\begin{figure}
\vskip  0cm
\centerline{\epsfig{file=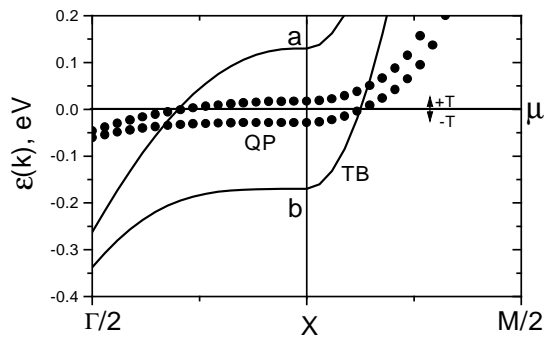,width=7cm,height=4.3cm}}
\vskip 0.2cm
\caption{ Energy spectrum for TB and QP near X-point.
Arrows indicate the room-temperature scale,
 $\Gamma /2 \equiv (\pi/2,0)$,
 $M/2 \equiv (\pi,\pi/2)$
($\delta$ =0.35, T=150 K).
}
\label{Ekqp}
\end{figure}

In order to show more clearly the formation of the flat
quasiparticle band in the bi-layer system we plot in the Fig.\ref{Ekqp}
the TB and QP spectrum for $\delta =0.35$ where
the Fermi level is approximately
between bonding and antibonding bands at X. While the
Fermi-level crossing-points are nearly
the same in the TB and QP bands,  the bonding
and antibonding QP bands are ``pinned'' to the Fermi level within room
temperature scale. In other words, the system has increased the spin and charge
fluctuations by pinning the saddle points to the Fermi level and
forming the
extended van-Hove singularities instead of the standard renormalization of the
spectrum only in the small region near the Fermi surface.

\begin{figure}
\vskip  0cm
\centerline{\epsfig{file=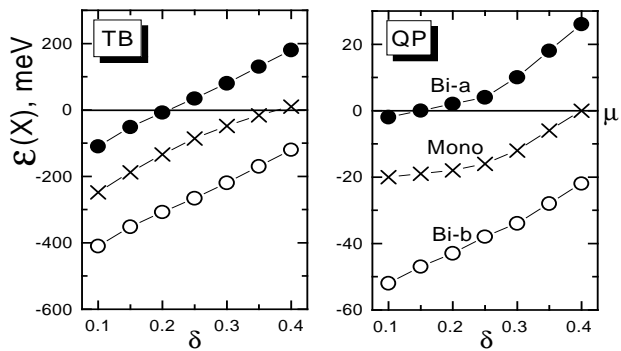,width=8cm,height=4.5cm}}
\vskip 0.2cm
\caption{ Position of the saddle point at X relative to
Fermi level for the TB and QP bands. Note the different scales.}
\label{pinqp}
\end{figure}

We have analyzed the effect of the saddle-points pinning to chemical potential
for the monolayer ($t_{\perp}=0$) and the bilayer models.
The energy position of the saddle-point at X with respect to the Fermi
level for the TB- and QP-spectrum as functions of the hole doping shown in
the Fig.\ref{pinqp}. The antibonding QP-band for the bi-layer model (Bi-a)
stays within the $\pm 5$ meV for large hole concentration range
$0.1 < \delta < 0.3$ per CuO$_2$. Such a pinning effect is less pronounced
in the monolayer (Fig.\ref{pinqp}).

The many-body reduction of the interlayer splitting also increases the
quasiparticle density of states in the bi-layer and could help the
formation of ``three-dimensional'' superconducting state.
We find the superconducting transition temperature by solving the
linearized Eliashberg equation for the singlet order
parameters in the bilayer model
\cite{Scalap-PRB}.
The critical temperature is defined via the maximum eigenvalue:
$\lambda _{\max }(T_c)>1$ of the following superconducting kernel:

$$
R_{k'k}=-\left(
\begin{array}{cc}
V_{+}^{sc}G_{+}(k)G_{+}(-k) & V_{-}^{sc}G_{+}(k)G_{+}(-k) \\
V_{-}^{sc}G_{-}(k)G_{-}(-k) & V_{+}^{sc}G_{-}(k)G_{-}(-k)
\end{array}
\right) ,
$$
where
$V_{\pm }^{sc}=V_{\pm }^s(k'-k)-V_{\pm }^c(k'-k)+U$.
We find that the superconducting
d$_{x^2-y^2}$ state, which is symmetrical with respect to the
mirror plane between the bi-layer, is indeed quite stable.
The transition temperature for
$\delta = 0.2$ is about 90 K (or T$_c$=0.015t), which is
almost the same as for single layer (within one percent).
The theoretical superconducting temperature for a LSCO-model\cite{MontScal}
with a nested Fermi surface is a bit higher.
If one takes into account the large order-parameter fluctuations
which are expected to reduce  T$_c$ in the single layer\cite{Lee},
we can expect higher T$_c$'s in the bi-layer systems.

We have repeated the calculations for the YBCO eight-band TB-model\cite{dimpl}
and found that the "bifurcated" saddle points in the TB-spectrum
helps to create the extended anisotropic van-Hove singularity.
In the YBCO model interlayer splitting is reduced by only
a factor of two and could be in principle detected by ARPES measurements.

In conclusion, we have shown that the generic
feature of bi-layer as well as mono-layer cuprate
is the formation of extended van-Hove singularities
near the chemical potential on the scale of the temperature,
which occurs more readily in bi-layers.
This leads to large areas of the Brilluoin zone with nearly
constant occupation $\approx 0.5$, which can explain the anomalous
normal state properties and lead to enhanced d$_{x^2-y^2}$ superconductivity.

We acknowledge stimulating conversations with P.W. Anderson, K.B. Efetov,
V.A. Khodel and M.C. Schabel.

\end{multicols}

\begin{references}

\bibitem[*]{Urb} On leave from:
UIUC at Urbana-Champaign.

\bibitem{PWAscience}  S. Chakravarty, A. Sudb\o , P.W.Anderson, and S. Strong,
Science, {\bf 261}, 337 (1993)

\bibitem{Schen-rev}  Z.-X. Schen {\it et al}.,
Science, {\bf 267}, 343 (1995).

\bibitem{Campuzano-BIS}  H.Ding {\it et al}.,
 Phys. Rev. Lett. {\bf 74}, 2784 (1995);
 H.Ding, A.F. Bellman {\it et al}.,
 preprint.

\bibitem{tranq}  J.M. Tranquada {\it et al}., Phys. Rev. B {\bf 46}, 5561
(1992).

\bibitem{Kanazawa} O.K. Andersen {\it et al}.,
 Physica C{\bf 185-189}, 147 (1991).

\bibitem{dimpl}  O.K.Andersen  {\it et al}.,
Phys. Rev. B {\bf 49}, 4145 (1994).

\bibitem{ScalapinoVH}  N. Bulut {\it et al}.,
Phys. Rev. B {\bf 50}, 7215 (1994).

\bibitem{DagottoVH}  E. Dagotto {\it et al}.,
 Phys. Rev. Lett. {\bf 73}, 728 (1994).

\bibitem{HankeVH}  R.Preuss  {\it et al}.,
 Phys. Rev. Lett. {\bf 75}, 1344 (1995).

\bibitem{Khodel} V.A. Khodel {\it et al}., JETP Lett.
{\bf 51}, 553 (1990).

\bibitem{Nozieres} P. Nozieres, J. Phys. {\bf 2}, 443, 1992.

\bibitem{DzyalVH} I.E. Dzyaloschinskii, Sov. Phys. JETP {\bf 66}, 848, 1987.

\bibitem{HankeFLEX} R. Putz, R. Preuss {\it et al}., preprint.

\bibitem{Scalapino-rev}  D.J. Scalapino, Physics Reports. {\bf 251}, 1 (1994).

\bibitem{Scalap-PRB}  N.Bulut  {\it et al}.,
Phys. Rev. B {\bf 45}, 5577 (1992).

\bibitem{Scalettar}  R.T. Scalettar {\it et al}.,
Phys. Rev. B {\bf 50}, 13419 (1994).

\bibitem{OKAperp}  O.K.Andersen {\it et al}.,
J. Phys. Chem. Solids, (1995).

\bibitem{U-LDA} A.K. McMahan {\it et al}.,
Phys. Rev. B {\bf 38}, 6650 (1988).

\bibitem{FLEX}  N.E. Bickers {\it et al}., Annals of Physics, {\bf 193},
206 (1989).

\bibitem{Serene}  J.W. Serene {\it et al}., Phys. Rev. B {\bf 44},
3391 (1991).

\bibitem{Kampf} A.P. Kampf {\it et al}.,
Phys. Rev. B {\bf 42}, 7967 (1990).

\bibitem{QP-Berlin} M. Langer, J. Schmalian {\it et al}., preprint.

\bibitem{Randiera} M. Randeria
{\it et al}., Phys. Rev. Lett. {\bf 74}, 4951 (1995)

\bibitem{Gop}S. Gopalan {\it et al}., Phys. Rev. B{\bf 46}, 11798 (1992).

\bibitem{MontScal}  P. Monthoux {\it et al}., Phys. Rev. Lett. {\bf 72}%
, 1874 (1994).

\bibitem{Lee}  P.A. Lee and N. Read, Phys. Rev. Lett. {\bf 25}, 2691, 1987.

\end{references}
\end{document}